\newcommand{\tJ}{{\it t-J}\,}
\newcommand{\fT}{{\cal T}}

\def\Per{P}

\newcommand{\LR}{\left[}
\newcommand{\RR}{\right]}

\documentclass[multphys,vecphys]{svmult}
\usepackage{amsmath}
\usepackage{amssymb}
\usepackage{epsf}
\usepackage{epsfig}
\usepackage{makeidx}
\usepackage{graphicx}
\usepackage{citesort}
\usepackage{multicol}    

\makeindex             
\newcommand{\nc}{\newcommand}
\nc{\rnc}{\renewcommand}
\nc{\be}{\begin{equation}}
\nc{\ee}{\end{equation}}
\nc{\bea}{\begin{eqnarray}}
\nc{\eea}{\end{eqnarray}}
\nc{\nn}{\nonumber}
\nc{\ch}{\cosh}
\nc{\sh}{\sinh}
\nc{\Z}{\overline{Z}}
\def\ba{\overline a}
\def\bA{\overline A}
\def\rhs{r.h.s.}

\def\refeq#1{(\ref{#1})}
\def\s#1{{s({#1})}}
\def\d#1{{d({#1})}}
\def\a{a}
\def\A{A}
\def\i{{\rm i}}
\def\e{{\rm e}}
\def\ut{{\tau}}
\nc{\mfa}{{\mathfrak{a}}}
\nc{\mfab}{\overline{\mfa}}
\nc{\mfA}{{\mathfrak{A}}}
\nc{\mfAb}{\overline{\mfA}}
\nc{\db}{\displaybreak[0]\\}
\newcommand{\zweimat}[4]{\left(\begin{array}{cc}
#1 & #2\\
#3 & #4\\
\end{array}\right)}
\newcommand{\ket}[1]{\left|#1\right>}

\newcommand{\trans}{\mathcal{T}}

\numberwithin{equation}{section}

\begin{document}
%
\title*{Integrability of quantum chains: theory and applications to 
the spin-1/2 $XXZ$ chain}
\author{Andreas Kl\"umper
}
\institute{Theoretische Physik, Universit\"at Wuppertal,
Gau\ss -Str.~20, D-42097 Wuppertal, Germany \texttt{kluemper@physik.uni-wuppertal.de}}
\titlerunning{Integrability of quantum chains, the spin-1/2 $XXZ$ chain}
\maketitle

%
%
\begin{abstract}
In this contribution we review the theory of integrability of quantum systems
in one spatial dimension. We introduce the basic concepts such as the
Yang-Baxter equation, commuting currents, and the algebraic Bethe
ansatz. Quite extensively we present the treatment of integrable quantum
systems at finite temperature on the basis of a lattice path integral
formulation and a suitable transfer matrix approach (quantum transfer
matrix). The general method is carried out for the seminal model of the
spin-1/2 $XXZ$ chain for which thermodynamic properties like specific heat,
magnetic susceptibility and the finite temperature Drude weight of the thermal
conductivity are derived.\\

\end{abstract}
%

%
\section{Introduction}
Integrable quantum chains have continuously attracted attention, because
of the possibility of obtaining exact data for the spectrum and other physical
properties despite the truely interacting nature of the spins resp.~particles.
Important examples of these systems are
the Heisenberg model \cite{Tak71,Gaudin71}, $tJ$-
\cite{Schlott92,JutKluSuz97,SugaOkiji97} and Hubbard models \cite{KUO89,JKS98}
The computational basis for
the work on integrable quantum chains is the Bethe ansatz yielding a set of
coupled non-linear equations for 1-particle wave-numbers (Bethe ansatz roots).
Many studies of the Bethe ansatz equations were directed at the
ground-states of the considered systems and have revealed interesting non-Fermi
liquid properties such as algebraically decaying correlation functions with
non-integer exponents and low-lying excitations of different types, i.e.~spin
and charge with different velocities constituting so-called spin and charge
separation, see \cite{Schulz93,KoEsBo,Lieb95}.

A very curious situation arises in the context of the calculation of the
partition function from the spectrum of the Hamiltonian. Despite the validity
of the Bethe ansatz equations for all energy eigenvalues of the above
mentioned models the direct evaluation of the partition function is rather
difficult. In contrast to ideal quantum gases the eigenstates are not
explicitly known, the Bethe ansatz equations provide just implicit
descriptions that pose problems of their own kind. Yet, knowing the behaviour
of quantum chains at finite temperature is important for many reasons. As a
matter of fact, the strict ground-state is inaccessible due to the very
fundamentals of thermodynamics. Therefore the study of finite temperatures is
relevant for theoretical as well as experimental reasons. 

The purpose of this review is to introduce to a unified treatment of the
ground-state and the finite temperature properties of integrable quantum
chains. First, in Sect.~\ref{IntegrHam} we introduce the essential concepts of
quantum integrability. We show how to construct an infinite set of conserved
currents, i.e.~local operators commuting with the Hamiltonian. At the heart of
these constructions is the embedding of the Hamiltonian into a family of
commuting row-to-row transfer matrices of a certain classical model.  In
Sect.~\ref{QTM} we derive a lattice path integral representation for the
partition function of a rather large class of integrable Hamiltonians at
finite temperature. Here we also introduce a very efficient transfer matrix
method based on the quantum transfer matrix. In Sect.~\ref{BA} we review the
algebraic Bethe ansatz for the seminal model of the partially anisotropic
spin-1/2 Heisenberg chain ($XXZ$ chain) related to the classical six-vertex
model on a square lattice. The results of this section are the construction of
eigenstates and eigenvalues of the Hamiltonian as well as the quantum transfer
matrix of the system.  In Sect.~\ref{BAENLIE} we demonstrate how to transform
the large number of coupled Bethe ansatz equations into a simple finite set of
non-linear integral equations. These equations are studied numerically in
Sect.~\ref{Numer} where explicit results for the temperature dependence of
specific heat and susceptibility of the spin-1/2 $XXZ$ chain are
given. Finally, in Sect.~\ref{transport} we resume the line of reasoning
developed for the proof of integrability and study the thermal conductivity of
the $XXZ$ chain a topic that is of considerable current interest.

\section{Integrable exchange Hamiltonians}
\label{IntegrHam}
We begin with the definition and discussion of a general class of quantum chains
with nearest-neighbour interactions based on (graded) permutations.  Consider
a one-dimensional lattice with $L$ sites and periodic boundary conditions
imposed. A $q$-state spin variable $\alpha_{i}$ is assigned to each site $i$.
We generally consider the situation where each spin $\alpha$ has its own
grading, i.e.~statistics number $\epsilon_{\alpha}=(-1)^{\xi_{\alpha}}=\pm
1$. A spin $\alpha$ with $\epsilon_{\alpha}=+1$ ($\epsilon_{\alpha}=-1$) is
called bosonic (fermionic). The Hamiltonian of the ``permutation model'' can
be introduced as 
\be 
H=\sum_{i=1}^{L}\Per_{i,i+1}
\label{eq:UShamiltonian} 
\ee 
with the (graded) permutation operator $\Per_{i,i+1}$
\be
\Per_{i.i+1}\ket{\alpha_{1}\cdots\alpha_{i}\alpha_{i+1}\cdots\alpha_{L}}
=(-1)^{\xi_{\alpha_{i}\alpha_{i+1}}}
\ket{\alpha_{1}\cdots\alpha_{i+1}\alpha_{i}\cdots\alpha_{L}},
\ee
where $\xi_{\alpha_{i}\alpha_{i+1}}$ is $1$ if both $\alpha_{i}$ and 
$\alpha_{i+1}$ are fermionic, and $0$ otherwise.

Model (\ref{eq:UShamiltonian}) is shown to be exactly solvable on the basis of
the Yang-Baxter equation. Many well-known exactly solvable models are of type
(\ref{eq:UShamiltonian}), e.g. the spin-$1/2$ Heisenberg chain with $q=2$ and
$\epsilon_{1}=\epsilon_{2}=+1$, the free fermion model with $q=2$ and
$\epsilon_{1}=-\epsilon_{2}=+1$, and the supersymmetric \tJ model with $q=3$
and $\epsilon_{1}=-\epsilon_{2}=\epsilon_{3}=+1$.  If $m$ of the $\epsilon$'s
are $+1$ and $n (=q-m)$ are $-1$, for example
$\epsilon_{1}=\cdots=\epsilon_{m}=+1$, $\epsilon_{m+1}=\cdots\epsilon_{q}=-1$,
we call the model $(m,n)$-permutation model or just $(m,n)$-model.

Before sketching the proof of integrability of the general permutation model
we look closer at two important special cases.  The $(2,0)$-model
($q=2,\,\epsilon_{1}=\epsilon_{2}=+1$) is the spin-1/2 Heisenberg chain with 
Hamiltonian 
\be
H=2\sum_{i=1}^{L}{\vec S}_{i}\cdot{\vec S}_{i+1} + L/2,
\label{isoHeis}
\ee
in terms of $SU(2)$ spin-1/2 operators ${\vec S}$.

The $(2,1)$-model ($q=3,\,\epsilon_{1}=\epsilon_{2}=+1$ and $\epsilon_{3}=-1$)
is the supersymmetric $\tJ$ model with Hamiltonian (ignoring a trivial shift)
\begin{equation}
H=-t\sum_{j,\sigma}{\cal P} (c^\dagger_{j,\sigma}c_{j+1,\sigma}+
  c^\dagger_{j+1,\sigma}c_{j,\sigma}) {\cal P} +J\sum_j(S_j
  S_{j+1}-n_j n_{j+1}/4),
\label{hamilton}  
\end{equation}
with standard fermionic creation and annihilation operators $c^\dagger$ and
$c$, projector ${\cal{P}}=\prod_j(1-n_{j\uparrow}n_{j\downarrow})$ ensuring
that double occupancies of sites are forbidden, and $2t=J$ (with normalization
$t=1$). The supersymmetric $\tJ$ model was shown to be integrable
\cite{Suth75,Sch87} by the well-known Bethe ansatz \cite{Bethe31,Yang67}.  The
ground-state and excitation spectrum were investigated \cite{BarBlaOg91} and
critical exponents calculated by finite-size scaling and conformal field
theory studies \cite{KawaYang90,BarKlu95}. The thermodynamical properties were
studied in \cite{SugaOkiji97} by use of the thermodynamical Bethe ansatz (TBA)
and in \cite{JutKlu96,JKS97} by use of the quantum transfer matrix (QTM).

\begin{figure}[!htb]
  \begin{center}
    \leavevmode
     \includegraphics[width=0.35\textwidth]{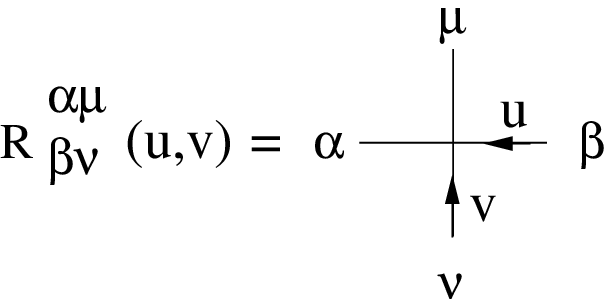}\hfill
    \caption{Graphical depiction of the fundamental $R$-matrix
       \refeq{eq:PS-model}. The Boltzmann weight assigned to each vertex
       configuration (configuration of spin variables around a vertex)
       corresponds to a matrix element of the matrix $R$.}
    \label{figR}
  \end{center}
\end{figure}

\begin{figure}[!htb]
  \begin{center} \leavevmode \includegraphics[width=0.5\textwidth]{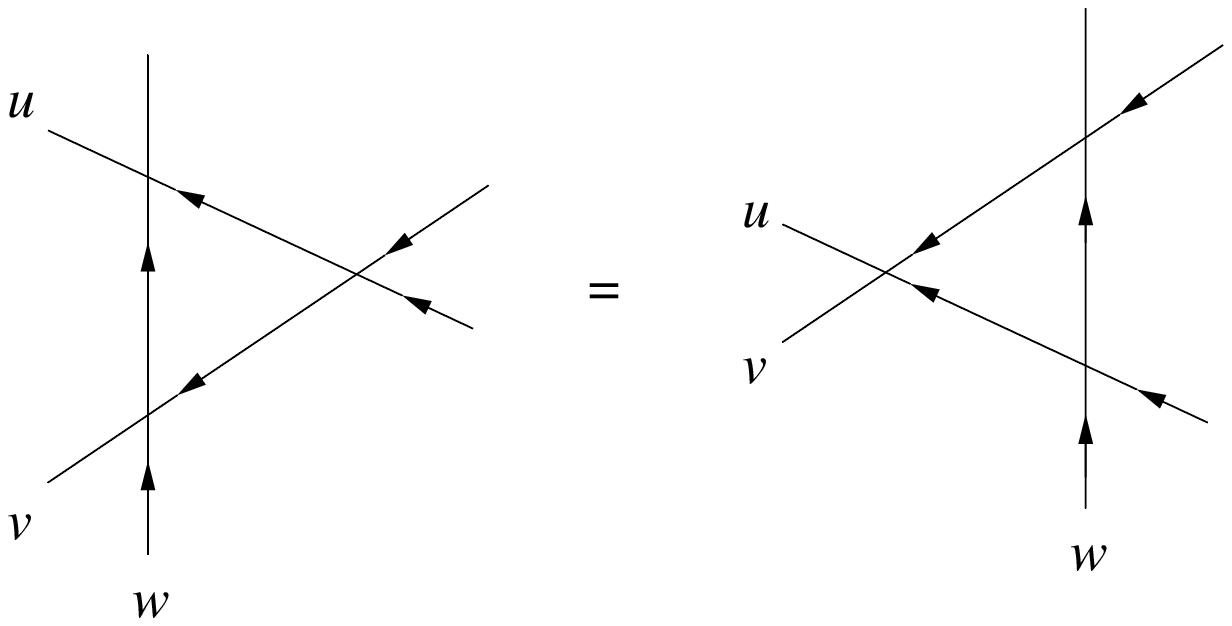}
    \caption{Depiction of the fundamental Yang-Baxter equation (YBE) for the
    $R$-matrix with the following graphical rules: Each bond carries a spin
    variable (ranging from 1 to $q$), each vertex (via Fig.~\ref{figR})
    corresponds to a local Boltzmann weight $R$ depending on the local spin
    configuration. The algebraic term corresponding to each graph of the
    equation is obtained by multiplying the Boltzmann weights corresponding to
    the (three) vertices and summing over spin variables on closed (inner)
    bonds. The values of the spin variables on open (outer) bonds are fixed
    and the assignment of these values is identical for both sides. The
    statement of the graphical equation is that both sides of the equation
    evaluate to the same expression for arbitrary, however identical
    assignment of spins to the open bonds on either side.  A consequence of
    the YBE is the commutativity of the transfer matrix with respect to the
    spectral parameter, see Sect.~\ref{QTM} and Fig.~\ref{fig:fig3b}.}
    \label{figYBEs} \end{center}
\end{figure}

For proving the integrability of the quantum system, a classical counterpart
is defined on a two-dimensional square lattice of $L\times N$ sites, where we
impose periodic boundary conditions throughout this paper.  We assume that
variables taking values $1,2,\cdots,q$ are assigned to the bonds of the
lattice. Boltzmann weights are associated with local vertex configurations
$\alpha$, $\beta$, $\mu$ and $\nu$ and are denoted by
$R^{\alpha\mu}_{\beta\nu}$, see Fig.~\ref{figR}.
The classical counterpart to (\ref{eq:UShamiltonian}) is the Perk-Schultz
model \cite{PerkSchultz81} with the following Boltzmann weights
\begin{equation}
R^{\alpha\mu}_{\beta\nu}(u,v)=\delta_{\alpha\nu}\delta_{\beta\mu}+
(u-v)\cdot (-1)^{\xi_{\alpha}\xi_{\mu}}\cdot\delta_{\alpha\beta}\delta_{\mu\nu},
\label{eq:PS-model}
\end{equation}
where $u$ and $v$ are freely adjustable ``interaction'' parameters assigned to
the entire horizontal and vertical lines intersecting in the particular vertex
under consideration.  These weights satisfy the Yang-Baxter equation which we
depict graphically in Fig.~\ref{figYBEs}.

In the remainder of this section we want to indicate: (i) how the classical
model is related to the quantum chain, and (ii) why the models are
integrable. These issues are best discussed in terms of the
transfer matrices of the classical model
\be
T^{\mu}_{\nu}(u) = \sum_{\{ \alpha\} }
                           \prod_{i=1}^{L}
                           R^{\alpha_{i}\mu_{i}}_{\alpha_{i+1}\nu_{i}}(u,v_i), 
\ee
where we consider the spectral parameters $v_i$ on the vertical bonds as
fixed, i.e.~independent of $u$. In dependence on the spectral parameter $u$
the object $T(u)$ represents a family of commuting matrices (the proof will be
given in a slightly more general setting in the subsequent section)
\be
T(v)T(w)=T(w)T(v),\qquad\hbox{for arbitrary}\ v, w.
\ee
Commutativity holds especially in the case of vanishing spectral parameters
$v_i=0$ on the vertical bonds, which case we refer to as the row-to-row
transfer matrix. For this we have the additional simple limiting behaviour
\bea
T(0)&=&\hbox{translation (shift) operator}=\e^{\i P},\\
\left. \frac d{du}\ln T(u)\right|_{u=0}&=& \hbox{Hamiltonian} = H, \label{eq:baxter}
\eea
also known as the Hamiltonian limit ($P$: momentum operator) \cite{Suth70,Bax82}.

Apparently, Hamiltonians obtained as members of commuting families of
operators possess (infinitely) many conserved quantities. Any element of the
family (or higher order derivative) commutes with $H$. For the case of
isotropic $SU(m,n)$-symmetric systems we have presented the typical proof of
integrability based on classical models satisfying the YBE. There are many
more models satisfying the YBE with different or reduced symmetries. A famous
example of a system with reduced symmetry is the partially anisotropic
spin-1/2 Heisenberg chain (also known as $XXZ$ chain). The Hamiltonian and
$R$-matrix corresponding to this will be given in the next section.

Above, we have only shown how to incorporate the momentum operator and the
Hamiltonian into the family of commuting operators $T(u)$.  As indicated, $\ln
T(u)$ is a generating function for conserved quantities
\be
\mathcal{J}^{(n)}=\left(\frac{\partial}{\partial u}\right)^n
\ln T(u) \Big|_{u=0},
\label{generator}
\ee
that appear to be sums of local operators (with or without physical
relevance). Quite generally, the current $\mathcal{J}^{(2)}$ is related to the
thermal current, whose conservation implies non-ballistic thermal transport,
see Sect.\ref{transport}.

Logically, the presentation of diagonalization of the Hamiltonian or, more
generally, of the row-to-row transfer matrix should follow directly after the
above discussion of integrability. We will see, however, that another class of
transfer matrices is worthwhile to be studied. This is a class of commuting
staggered transfer matrices (quantum transfer matrices) occurring in the study
of the thermodynamics of the quantum system at finite temperature (next
section). We therefore postpone the discussion of the algebraic Bethe ansatz
for both classes of transfer matrices to Sect.~\ref{BA}.

%
\section{Lattice path integral and quantum transfer matrix}
\label{QTM}
At the beginning of this section we want to comment on the various techniques
developed for the study of thermodynamics of integrable systems at the example
of the simplest one, namely the spin-1/2 Heisenberg chain. 

The thermodynamics of the Heisenberg chain was studied in
\cite{Tak71,Gaudin71,Takahashi71b} by an elaborate version of the method used
in \cite{YangYang69}. The macro-state for a given temperature $T$ is described
by a set of density functions formulated for the Bethe ansatz roots satisfying
integral equations obtained from the Bethe ansatz equations \refeq{BAgeneral}.
In terms of the density functions expressions for the energy and the entropy
are derived. The minimization of the free energy functional yields what is
nowadays known as the Thermodynamical Bethe Ansatz (TBA).

There are two ``loose ends'' in the sketched procedure. First, the
description of the spectrum of the Heisenberg model is built on the so-called
``string hypothesis'' according to which admissible Bethe ansatz patterns of
roots are built from regular building blocks.  This hypothesis was criticized
a number of times and led to activities providing alternative approaches to the
finite temperature properties.

The second ``loose end'' within TBA concerns the definition of the entropy
functional. In \cite{Tak71,Gaudin71,Takahashi71b,YangYang69} the entropy is
obtained from a combinatorial evaluation of the number of micro-states
compatible with a given set of density functions of roots.  As such it is a
lower bound to the total number of micro-states falling into a certain energy
interval. However, this procedure may be viewed as a kind of saddle point
evaluation in the highly dimensional subspace of all configurations falling
into the given energy interval.  Hence, the result is correct in the
thermodynamic limit and the ``second loose end'' can actually be tied up.
Interestingly, the ``second loose end'' of the TBA approach was motivation for
a ``direct'' evaluation \cite{KatWad02} of the partition function of
integrable quantum chains.  A straightforward (though involved) calculation
leads to the single non-linear integral equation of
\cite{Tak00}.

In this section we want to introduce the approach to thermodynamics of
integrable quantum chains that we believe is the most efficient one, namely
the ``quantum transfer matrix'' (QTM) approach.  The central idea of this
technique is a lattice path-integral formulation of the partition function of
the Hamiltonian and the definition of a suitable transfer
matrix\cite{Koma87,SuzIno87,Bariev82,TruSch83,SuzAkuWad90,SuzNagWad92,Tak91,Klu92,Klu93}.

In order to deal with the thermodynamics in the canonical ensemble we have to
deal with exponentials of the Hamiltonian $H$. These operators are obtained
from the row-to-row transfer matrix $T(u)$ of the classical model in the
Hamiltonian limit (small spectral parameter $u$) \refeq{eq:baxter}
\begin{equation}
T(u) = \e^{\I P+u H+O(u^2)},
\end{equation}
with $P$ denoting the momentum operator.  

\begin{figure}[!htb]
  \begin{center}
    \leavevmode
     \includegraphics[width=0.8\textwidth]{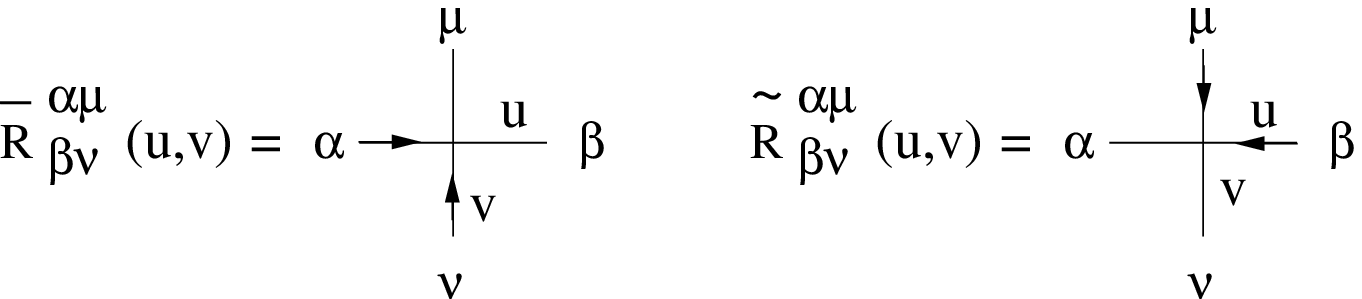}\hfill
    \caption{Graphical depiction of the $\overline R$ and $\widetilde R$
       matrices (\refeq{Rbar} and \refeq{Rtilde}) associated with $R$. The
       Boltzmann weights assigned to the vertex configurations $\overline R$
       and $\widetilde R$ correspond to the Boltzmann weights of the rotated
       fundamental vertices $R$.}
    \label{figRbt}
  \end{center}
\end{figure}

\begin{figure}[!htb]
  \begin{center}
    \leavevmode
      \includegraphics[width=0.8\textwidth]{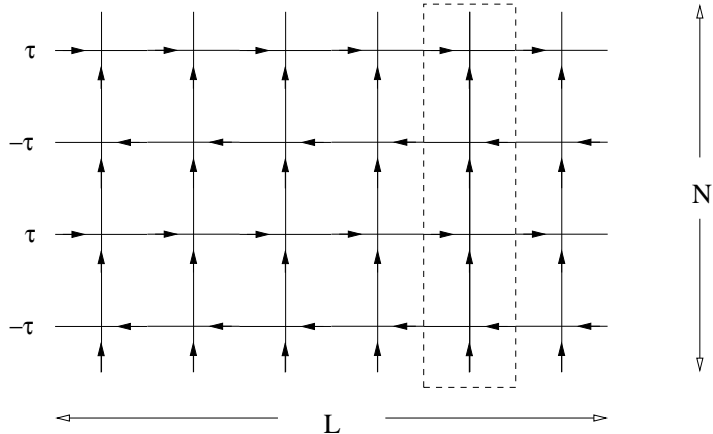}
    \caption{Illustration of the two-dimensional classical model onto which
      the quantum chain at finite temperature is mapped. The square lattice
      has width $L$ identical to the chain length, and height identical to the
      Trotter number $N$.  The alternating rows of the lattice correspond to
      the transfer matrices $T(-\ut)$ and $\overline{T}(-\ut)$, $\ut=\beta/N$.
      The column-to-column transfer matrix $T^{{\rm QTM}}$ (quantum transfer
      matrix) is of particular importance to the treatment of the
      thermodynamic limit. The arrows placed on the bonds indicate the type of
      local Boltzmann weights, i.e.~$R$ and $\overline R$-matrices alternating
      from row to row. (The arrows indicate the type of Boltzmann weight, they
      do not denote local dynamical degrees of freedom.)}
    \label{fig:fig2Gen}
  \end{center}
\end{figure}

The main idea of the quantum transfer matrix (QTM) method at finite
temperature is as simple as follows (for details the reader is referred to the
papers \cite{Klu92,Klu93}).  First, let us define a new set of vertex weights
${\overline R}$ by rotating $R$ by 90 degrees as
\be 
{\overline R}^{\alpha\mu}_{\beta\nu}(u,v)=R^{\mu\beta}_{\nu\alpha}(v,u),\label{Rbar}
\ee
see Fig.~\ref{figRbt}. We further introduce an ``adjoint'' transfer matrix
${\overline T}(u)$ as a product of ${\overline R}(-u,0)$ \cite{Klu92,JKS98}
with Hamiltonian limit
\begin{equation}
\overline{T}(u) = \e^{-\I P+u H+O(u^2)}.
\end{equation}
With these settings the partition function $Z_L$ of the quantum chain of
length $L$ at finite temperature $T$ reads
\begin{equation}
Z_L =  \hbox{Tr}\,\e^{-\beta H} = 
      \lim_{N\rightarrow \infty} Z_{L,N}
\label{partL}
\end{equation}
where $\beta=1/T$ and $Z_{L,N}$ is defined by
\begin{equation}
Z_{L,N}:=\hbox{ Tr } \left[T(-\ut)
  \overline{T}(-\ut)\right]^{N/2},\qquad\ut:=\frac\beta N.
\label{Trotter}
\end{equation}
The \rhs of this equation can be interpreted as the partition function of a
staggered vertex model with alternating rows corresponding to the transfer
matrices ${T}(-\ut)$ and $\overline{T}(-\ut)$, see
Fig.~\ref{fig:fig2Gen}. 
We are free to evaluate the partition function of this classical model by
adopting a different choice of transfer direction. A particularly useful
choice is based on the transfer direction along the chain and on the
corresponding transfer matrix $T^{{\rm QTM}}$ which is defined for the columns
of the lattice.  The partition function of the quantum chain at temperature
$1/\beta$ is given by
\begin{equation}
Z_{L,N}=\hbox{ Tr } \left(T^{{\rm QTM}}\right)^L.
\label{Trotter2}
\end{equation}
In the remainder of this paper we will refer to $T^{{\rm QTM}}$ as the
``quantum transfer matrix'' of the quantum spin chain, because $T^{{\rm QTM}}$
is the closest analogue to the transfer matrix of a classical spin chain. Due
to this analogy the free energy $f$ per lattice site is given just by the
largest eigenvalue $\Lambda_{{\rm max}}$ of the QTM
\begin{equation}
f= -k_B T \lim_{N\rightarrow \infty}\log\Lambda_{{\rm max}}.
\label{freeEnerEig}
\end{equation}
Note that the eigenvalue depends on the argument $\ut={\beta}/{N}$
which vanishes in the limit $N\rightarrow \infty$ requiring
a sophisticated treatment.

The QTM as defined above is actually a member of a commuting family of
matrices $\fT^{\rm QTM}$ defined by
\be
\LR T^{\rm QTM}\RR^{\mu}_{\nu}(v)=\sum_{\alpha}\prod_{j=1}^{N/2}
R^{\alpha_{2j-1}\mu_{2j-1}}_{\alpha_{2j}\nu_{2j-1}}(v,\ut)
{\widetilde R}^{\alpha_{2j}\mu_{2j}}_{\alpha_{2j+1}\nu_{2j}}(v,-\ut), 
\label{QTMfam}
\ee
where we introduced yet another vertex weight ${\tilde R}$ as a
rotation of $R$ by -90 degrees
\be
{\widetilde R}^{\alpha\mu}_{\beta\nu}(u,v)=R_{\mu\beta}^{\nu\alpha}(v,u).
\label{Rtilde}
\ee
Here we have introduced a spectral parameter $v$ such that $\fT_{\rm QTM}(v)$
is a commuting family of matrices generated by $v$. A proof of this 
\begin{figure}[!htb]
  \begin{center}
    \leavevmode
    \includegraphics[width=0.5\textwidth]{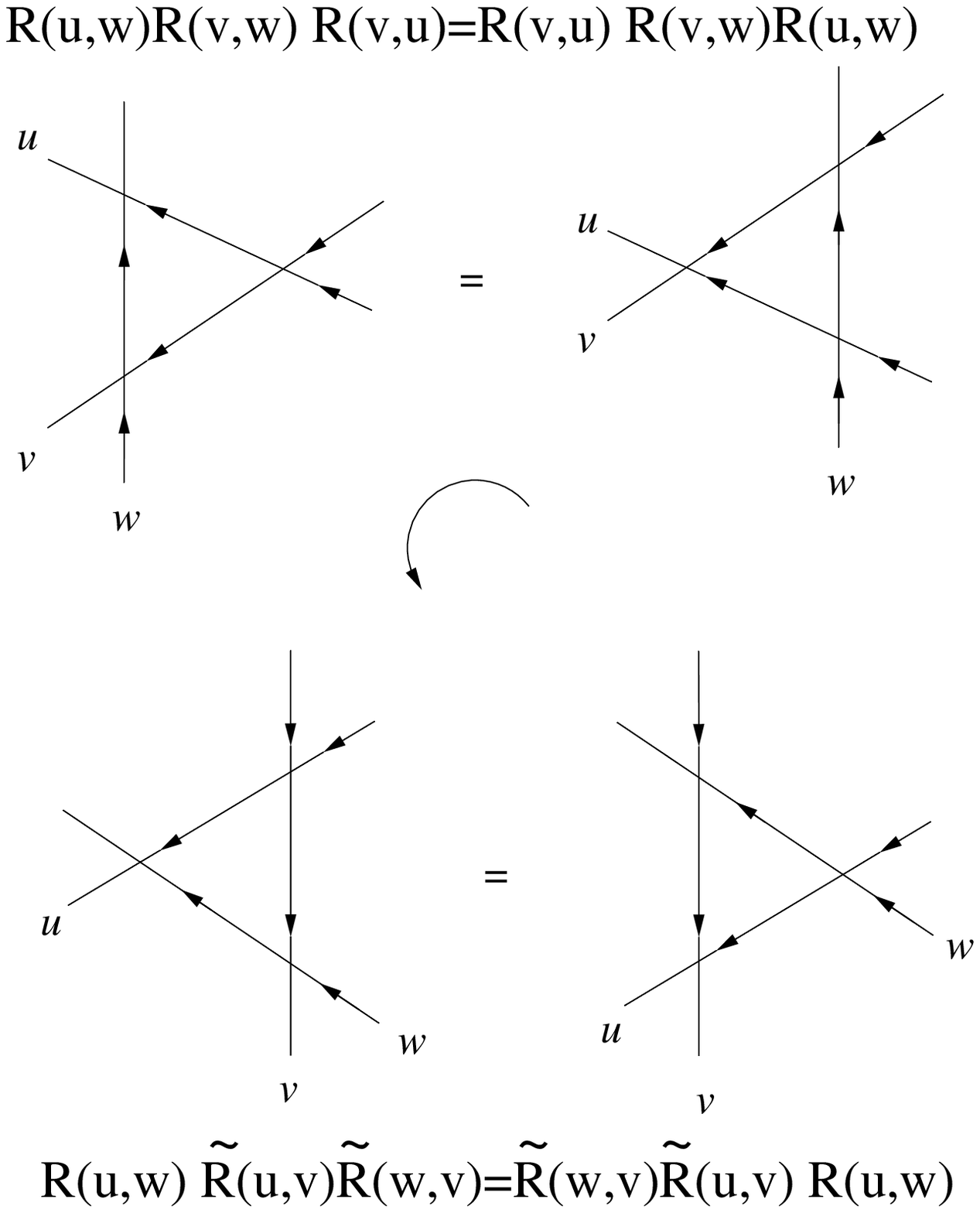}
    \caption{Graphical derivation of the Yang-Baxter equation for $\widetilde
    R$ matrices. In the upper row, the fundamental YBE for $R$ matrices is
    shown. The YBE for $\widetilde R$ matrices is obtained through
    rotation. Note that the intertwiner for $R$ vertices is identical to the
    intertwiner for $\widetilde R$ vertices.}
    \label{fig:fig3}
  \end{center}
\end{figure}
consists of two steps. First, we observe that $\widetilde R$ matrices and $R$
matrices share the same intertwiner, i.e.~the order of multiplication of two
$\widetilde R$ matrices is interchanged by the same $R$ matrix as in the case
of the fundamental YBE, see Fig.~\ref{fig:fig3}.
\begin{figure}[!htb]
  \begin{center}
    \leavevmode
    \includegraphics[width=0.5\textwidth]{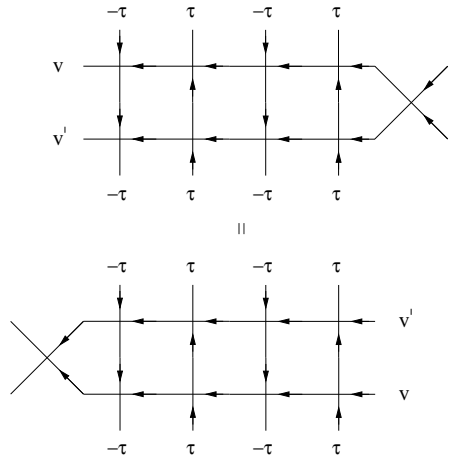}
    \caption{Railroad proof for the commutativity of transfer matrices: The
intertwiner is pulled through from right to left by a successive application
of the YBE.}
    \label{fig:fig3b}
  \end{center}
\end{figure}
As $R$ and $\widetilde R$ matrices share the same intertwiner, the transfer
matrix obtained from products of arbitrary sequences of $R$ and $\widetilde R$
with same spectral parameter (say $v$) on the continuous line constitutes a
family of commuting matrices (spanned by $v$). The proof of this statement is
graphically depicted in Fig.~\ref{fig:fig3b} for the case of the matrix $T^{\rm
QTM}(v)$ \refeq{QTMfam}.  In other words, $T^{\rm QTM}(v)$ is integrable
which allows us to diagonalize $T_{\rm QTM}(v)$. The final results, of
course, are physically interesting just for $v=0$ as the physically meaningful
QTM is identical to $T^{\rm QTM}(0)$.

The main difference to the transfer matrix treatment of classical spin chains
is the infinite dimensionality of the space in which $T^{{\rm QTM}}$ is
living (for ${N\rightarrow \infty}$).  In formulating \refeq{freeEnerEig} we
have implicitly employed the interchangeability of the two limits ($ L, N
\rightarrow \infty$) and the existence of a gap between the largest and the
next-largest eigenvalues of $T^{{\rm QTM}}$ for finite temperature
\cite{SuIn87,SAW90}.

The next-leading eigenvalues give the exponential correlation lengths $\xi$ of
the equal time correlators at finite temperature
\be
\frac{1}{\xi}=\lim_{N\rightarrow
  \infty}\ln\left|\frac{\Lambda_{{\rm max}}}{\Lambda}\right|.
\label{length}
\ee

Lastly we want to comment on the study of thermodynamics of the quantum
chain in the presence of an external magnetic field $h$ coupling to the
spin $S=\sum_{j=1}^LS_j$, where $S_j$ denotes a certain component of the 
$j$th spin, for instance $S_j^z$. This, of course, changes \refeq{Trotter}
only trivially
\begin{equation}
Z_{L,N}:=\hbox{ Tr } \left\{\left[T(-\ut) \overline{T}(-\ut)\right]^{N/2}
\cdot\e^{\beta h S}\right\}.
\end{equation}
On the lattice, the equivalent two-dimensional model is modified in a simple
way by a horizontal seam. Each vertical bond of this seam carries an
individual Boltzmann weight $\e^{\pm\beta h/2}$ if $S_j=\pm1/2$ which indeed
describes the action of the operator
\be
\e^{\beta h S}=\prod_{j=1}^{L}\e^{\beta h S_j}.
\ee
Consequently, the QTM is modified by an $h$ dependent 
boundary condition. It is essential that these modifications can still be
treated exactly as the additional operators acting on the bonds belong
to the group symmetries of the model.

We like to close this section with some notes on the relation of the two
apparently different approaches, the combinatorial TBA and the operator-based
QTM. In fact, these methods are not at all independent! In the latter approach
there are several quite different ways of analysis of the eigenvalues of the
QTM.  In the standard (and most economical) way, see below, a set of just two
coupled non-linear integral equations (NLIE) is derived \cite{Klu92,Klu93}.
Alternatively, an approach based on the ``fusion hierarchy'' leads to a set of
(generically) infinitely many NLIEs \cite{Klu92,KSS98} that are identical to
the TBA equations though completely different reasoning has been applied!

Very recently \cite{Tak00}, yet another formulation of the thermodynamics of
the Heisenberg chain has been developed. At the heart of this formulation is
just one NLIE with a structure very different from that of the two sets of
NLIEs discussed so far. Nevertheless, this new equation
has been derived from the ``old'' NLIEs \cite{Tak00,TakShirKl01} and is
certainly an equivalent formulation. In the first applications of the new
NLIE, numerical calculations of the free energy have been performed with
excellent agreement with the older TBA and QTM results. Also, analytical high
temperature expansions up to order 100 have been carried out on this basis.

\section{Bethe ansatz equations for the spin-1/2 $XXZ$ chain}
\label{BA}
In the following we consider the anisotropic Heisenberg chain (slightly generalizing
\refeq{isoHeis}) with Hamiltonian $H_L$
\begin{equation}
H=2\sum_{j=1}^L[S^x_j S^x_{j+1}+ S^y_j S^y_{j+1}+\Delta S^z_j S^z_{j+1}]
\label{anisoHeis}
\end{equation}
with periodic boundary conditions on a chain of length $L$. Apparently, for
$\Delta=+ 1$ the system specializes to the isotropic antiferromagnetic
Heisenberg chain, for $\Delta=- 1$ (and applying a simple unitary
transformation) the system reduces to the isotropic ferromagnetic case.

The classical counterpart of the $XXZ$ chain is the six-vertex model.  For our
purposes the following parameterization of the Boltzmann weights is useful 
\be
a(w)=1,\quad b(w)=\frac{\sin(\gamma w/2)}{{\sin(\gamma w/2+\gamma)}},\quad
c(w)=\frac{\sin\gamma}{\sin(\gamma w/2+\gamma)}.
\label{6VMparametr}
\ee
\begin{figure}[!htb]
  \begin{center}
    \leavevmode
      \includegraphics[width=0.6\textwidth]{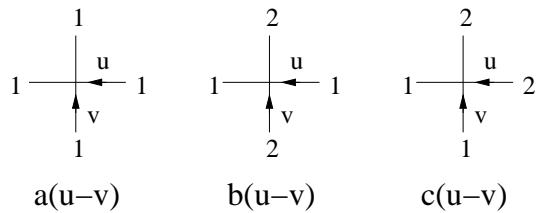}
    \caption{Allowed vertices of the six-vertex models with
    corresponding Boltzmann weights $a$, $b$, $c$. The remaining
    vertices are obtained by an exchange of bond-states $1\leftrightarrow 2$.}
    \label{fig:fig4b}
  \end{center}
\end{figure}
All relations between Hamiltonian of the quantum system, row-to-row transfer
matrix of the classical model and the QTM as introduced in the previous
section are also valid in the present case if $\Delta$ and $\gamma$ are
related by $\Delta=\cos\gamma$. The only modification concerns
\refeq{eq:baxter} as here the relation of Hamiltonian \refeq{anisoHeis} and
the logarithmic derivative of the row-to-row transfer matrix for the weights
\refeq{6VMparametr} aquires a normalization factor
\be
H=2\frac{\sin\gamma}\gamma\frac d{du}\ln T(u)\big|_{u=0}.\label{HamXXZ}
\ee

\noindent
{\it Monodromy matrix}\\ 
Our aim is to diagonalize the row-to-row transfer matrix and the QTM by means
of the algebraic Bethe ansatz.  We first review some notation and basic
properties of $R$-matrices (as collections of local Boltzmann weights) and the
so-called $L$-matrix. The elements of the $L$-matrix at site $j$ are
operators acting in the local Hilbert space $h_j$ (for the six-vertex model
isomorphic to $\simeq{\mathbb C}^2$).  The $L$-matrix' element in row $\alpha$
and column $\beta$ is given in terms of the $R$ matrix

\begin{equation}
{L_j}^\alpha_\beta(w)=R^{\alpha \mu}_{\beta \nu}(w){e_j}_\nu^\mu,
\label{LfromR}
\end{equation}
where $e_\nu^\mu$ is a matrix with only non-vanishing entry 1 in
row $\mu$ and column $\nu$. This reads explicitly for the six-vertex model
\begin{equation}
L_i=\zweimat{\frac{a+b}{2}+\frac{a-b}{2}\sigma_i^z}{c\sigma_i^+}
{c\sigma_i^-}{\frac{a+b}{2} -\frac{a-b}{2}\sigma_i^z}
\label{6VMLoperator}
\end{equation}
with $a,b,c$ given in \refeq{6VMparametr}.\footnote{As a reminder:
$\sigma^+=\sigma^x+i\sigma^y$ and $\sigma^-=\sigma^x-i\sigma^y$, and states
$\ket{+}$, $\ket{-}$ correspond to $\ket{1}$, $\ket{2}$.}  The
so-called monodromy-matrix is defined as a product of all $L$-matrices
on consecutive sites
\begin{equation} 
\mathcal{T}=L_1\cdots L_N=\zweimat{A}{B}{C}{D}.
\end{equation}
In essence, the $(\alpha,\beta)$-element of the monodromy matrix is the
transfer matrix of a system with fixed boundary spins $\alpha$ on the left end
and $\beta$ on the right end of a row.\\

\noindent
{\it Algebraic Bethe ansatz}\\ 
 The procedure of diagonalization can be decribed as follows:

\begin{itemize}
\item First we search for a pseudo-vacuum state (reference state)
$\ket{\Omega}$ that is a simple eigenstate of the operator-valued diagonal
entries $A$ and $D$ of the monodromy matrix ${\cal T}$ (and hence an
eigenstate of the transfer matrix $T=A+D$). The lower-left entry $C$ of the
monodromy matrix applied to $\ket{\Omega}$ yields zero, the upper-right entry
$B$ yields new non-vanishing states. Hence $C$ and $B$ play the role of
annihilation and creation operators.\\

\item From the Yang-Baxter equations a quadratic algebra of commutation
relations for the entries of the monodromy matrix (notably for $A$, $D$, and
$B$) is derived. By use of these relations algebraic expressions of the
eigenstates and rather explicit expressions for the eigenvalues are derived.
\end{itemize}

It turns out that the row-to-row transfer matrix and the QTM can be treated in
parallel as the quadratic algebra of the entries of the monodromy matrix is
identical for both cases. The only difference lies in the different reference
states and the different ``vacuum expectation values''. We therefore focus
our presentation on the slightly simpler case of the row-to-row transfer
matrix and describe the necessary modifications for the QTM in
(\ref{replacementab},\ref{replacementab2}).\\

\noindent
{\it Reference state}\\ 
Provided we find local states $\ket{\omega_i}$ that
are eigenstates of the diagonal entries of $L_i$ and are annihilated by the
left-lower entry, then $\ket{\Omega}$ may be taken as the product of these
local states. A glance to \refeq{6VMLoperator} shows that this is possible just
with $\ket{\omega_i}=\ket{2}_i$
\begin{equation}
\ket{\Omega}=\bigotimes_i^N \ket{2}_i.
\end{equation}

The monodromy matrix $\cal{T}$ applied to $\ket{\Omega}$ yields an upper
triangular $2 \times 2$ matrix of states
\begin{equation}
\trans\ket{\Omega}=\left(\begin{array}{cc}
a^N \ket{\Omega} & B\ket{\Omega}\\
0 & b^N \ket{\Omega}\\
\end{array}\right)
\end{equation} 
or explicitly
\begin{equation}
A\ket{\Omega}=a^N\ket{\Omega},\quad
D\ket{\Omega}=b^N\ket{\Omega},\quad
T\ket{\Omega}=(a^N+b^N)\ket{\Omega}.
\end{equation} 
Therefore, $\ket{\Omega}$ is an eigenstate of $T$.\\ 

\noindent
{\it Quadratic algebra of operators $A$, $B$, and $D$}\\
We intend to use the operator $B$ as creation operator for excitations, i.e.~we demand
that the new state $\ket{\Omega_1(v)}:=B(v)\ket{\Omega}$ (``one-particle
state'') be an eigenstate of $T(u)=A(u)+D(u)$. What we need to know is
the operator algebra for interchanging $B(v)$ with
$A(u)$ and $D(u)$. This algebra can be obtained from the YBE, for
a graphical representation see Fig.~\ref{yb}.
\begin{figure}[ht]
\begin{center}
\vspace{0.5cm}
\input{Yangbax.pstex_t}
\caption{Graphical depiction of the YBE.} 
\label{yb}
\end{center}
\end{figure}
By fixing the exterior spins on the horizontal bonds we obtain all the
commutators we need for interchanging the operators. We begin with the
relation of any two $B$-operators illustrated in Fig.~\ref{bop}
actually implying commutation
\begin{equation}
[B(u),B(v)]=0.
\end{equation}
\begin{figure}[ht]
\vspace{0.5cm}
\begin{center}
\input{Bop2.pstex_t}
\caption{Quadratic relation of $B$-operators obtained from fixing the spins on
the left and right open bonds in Fig.~\ref{yb}. Note that the summation over
the spin variables on the closed bonds between $R$ and $\cal T$ matrices
reduces to just one non-vanishing term on each side of the equation.}
\label{bop}
\end{center}
\end{figure}

Now we look at products of $A$ and $B$ and again use a graphical
representation shown in Fig.~\ref{aop}.
We algebraically find
\begin{equation}
A(u)B(v)=\frac{a(v-u)}{b(v-u)}B(v)A(u)-\frac{c(v-u)}{b(v-u)}B(u)A(v).\label{ABcomm}
\end{equation}
\begin{figure}[ht]
\vspace{0.5cm}
\begin{center}
\input{Aop2.pstex_t}
\caption{Bilinear relation of $B$ and $A$-operators obtained from fixing the spins on
the left and right open bonds in Fig.~\ref{yb}. Note that the summation over
the spin variables on the closed bonds between $R$ and $\cal T$ matrices
reduces to two non-vanishing terms on the left side of the equation and to
one non-vanishing term on the right side of the equation.}
\label{aop}
\end{center}
\end{figure}

In a similar same way, illustrated in Fig.~\ref{dop}, we obtain (after
interchanging $u\leftrightarrow v$)
\begin{equation}
D(u)B(v)=\frac{a(u-v)}{b(u-v)}B(v)D(u)-\frac{c(u-v)}{b(u-v)}B(u)D(v).\label{DBcomm}
\end{equation}
\begin{figure}[ht]
\vspace{0.5cm}
\begin{center}
\input{Dop2.pstex_t}
\caption{Bilinear relation of $B$ and $D$-operators obtained in analogy to the
above cases.} \label{dop}
\end{center}
\end{figure}

Now we have available all necessary relations for evaluating the application
of $T(u)$ to $\ket{\Omega_1}$. By use of the commutation relations we get
\begin{eqnarray}
\label{ewg}
T(u)\ket{\Omega_1(v)}&=&\left[\alpha(u)\frac{a(v-u)}{b(v-u)}
+\beta(u)\frac{a(u-v)}{b(u-v)}\right]\ket{\Omega_1(v)}\nonumber\\
&-&\left[\alpha(v)\frac{c(v-u)}{b(v-u)}
+\beta(v)\frac{c(u-v)}{b(u-v)}\right]B(u)\ket{\Omega},
\end{eqnarray} 
where we have used the abbreviations $\alpha=a^N,\,\, \beta=b^N$.  The terms
in the first (second) line of the \rhs of \refeq{ewg} are called ``wanted
terms'' (``unwanted terms'' ) and arise due to the first (second) terms on the
\rhs of (\ref{ABcomm},\ref{DBcomm}). In order that $\ket{\Omega_1(v)}$ be an
eigenstate of the transfer matrix the second term in (\ref{ewg}) has to vanish
and the first one to give the eigenvalue\footnote{For the sake of transparency
of the algebraic structure we do not use the fact that by our convention the
Boltzmann weight $a$ is identical to 1.}
\begin{eqnarray}
\label{ewgg}
\frac{\alpha(v)}{\beta(v)}&=&-\frac{c(u-v)b(v-u)}{c(v-u)b(u-v)}=1,\\
\Lambda(u)&=&\alpha(u)\frac{a(v-u)}{b(v-u)}+\beta(u)\frac{a(u-v)}{b(u-v)}. 
\end{eqnarray}  
We can generalize this to any $n$-particle state. The argument is quite the same as
for just one excitation. We look at the following state
\begin{equation}
\ket{\Omega({v_i})}=\prod_{i=1}^n B(v_i)\ket{\Omega}
\end{equation} 
where the numbers ${v_i}$ will be referred to as Bethe ansatz roots. Demanding that this
state be an eigenstate of $T(u)$ we get after successive application of the
commutation rules the following set of equations. From the two ``wanted
terms'' the eigenvalue is read off
\be
\Lambda(u)=\alpha(u)\prod_{j=1}^n\frac{a(v_j-u)}{b(v_j-u)}
+\beta(u)\prod_{j=1}^n\frac{a(u-v_j)}{b(u-v_j)},\label{BAgeneral1}
\ee
and vanishing of the ``unwanted terms'' yields
\be
\frac{\alpha(v_i)}{\beta(v_i)}=\prod\limits_{j (\not=i)}^n
\frac{b(v_j-v_i)}{b(v_i-v_j)},\quad\mbox{for}\;i=1,...,n.\label{BAgeneral}
\ee
The last constraints are nothing but the famous Bethe ansatz equations. We
like to note that we would have obtained the same set of equations by
demanding that the function on the \rhs of \refeq{BAgeneral1} be analytic in
the whole complex plane
\begin{equation}
\label{analytic}
\mbox{Res}\,\, \Lambda (u=v_i)=0 \qquad\forall\; i.
\end{equation} 

In the case of a general transfer matrix
${T}\left(v;\{u^{(1)}_k\};\{u^{(2)}_k\}\right)$ defined as a product of $N_1$
many vertices of type $R$ and $N_2$ many vertices of type $\widetilde R$ with
spectral parameter $v$ in the auxiliary space (corresponding to the continuous
line) and $u^{(1)}_k$, $u^{(2)}_k$ in the quantum spaces the eigenvalue
expression is very similar to
\refeq{BAgeneral1} and \refeq{BAgeneral}. The only change is the replacement
\bea
\alpha(u)&=&\e^{+\beta h/2} 
\prod_{k=1}^{N_1}a(u-u^{(1)}_k)\prod_{k=1}^{N_2}b(u^{(2)}_k-u),\label{replacementab}\\
\beta(u)&=&\e^{-\beta h/2} 
\prod_{k=1}^{N_1}b(u-u^{(1)}_k)\prod_{k=1}^{N_2}a(u^{(2)}_k-u),\label{replacementab2}
\eea
where we have also introduced $\exp(\pm\beta h/2)$ factors arising from twisted
boundary conditions for the transfer matrix as realized in the case of the
quantum transfer matrix for a system in a magnetic field $h$.

We may simplify the following presentation if we perform a rotation by
$\pi/2$ in the complex plane, i.e.~we introduce a function $\lambda$ by
$\lambda(v)=\Lambda(\I v)$ and for convenience we replace $v_j \rightarrow \I v_j$. 
The eigenvalue expression for any eigenvalue $\lambda(v)$ of the
transfer matrix $T$ reads (see also \cite{Bax82})
\be
\lambda(v)=\alpha(\I v)\prod_{j=1}^n\frac{a(\I v_j-\I v)}{b(\I
  v_j-\I v)}
+\beta(\I v)\prod_{j=1}^n\frac{a(\I v-\I v_j)}{b(\I v-\I v_j)}.\label{eigrow}
\ee
Of particular importance is the case of the quantum transfer matrix with
$N_1=N_2=N/2$ and all $u^{(1)}_k=\tau$, $u^{(2)}_k=-\tau$ with
$\tau=2\frac{\sin\gamma}\gamma\frac\beta N$, note the normalization
factor in \refeq{HamXXZ}.

\begin{figure}[!htb]
  \begin{center}
    \leavevmode
      \includegraphics[width=0.7\textwidth]{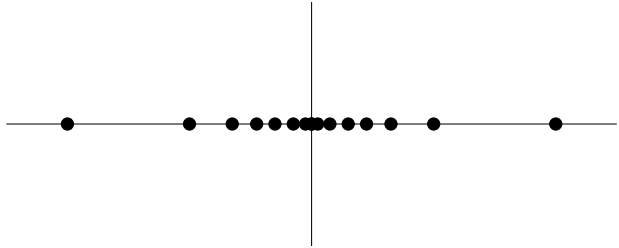}
    \caption{Sketch of the distribution of Bethe ansatz roots 
        $v_j$ for finite $N$. Note that the distribution remains 
        discrete in the limit of $N\to\infty$ for which the origin 
        turns into an accumulation point.}
    \label{fig:DistBAR}
  \end{center}
\end{figure}

Introducing the definition
\be
 \s{v}:=\sinh(\gamma v/2),
\ee
we obtain for the functions $b$, $\alpha$, $\beta$
\begin{eqnarray}
b(\I v)&=&\frac{\s{v}}{\s{v-2\I}},\cr
\alpha(\I v)&=&\e^{\beta h/2}\left[\frac{\s{v-\I\ut}}{\s{v-\I\ut+2\I}}\right]^{N/2},\quad
\beta(\I v)=\e^{-\beta h/2}\left[\frac{\s{v+\I\ut}}{\s{v+\I\ut-2\I}}\right]^{N/2}.
\end{eqnarray}
From \refeq{eigrow} we obtain for the function $\lambda(v)$
\begin{equation}
\lambda(v)=\frac{\lambda_1(v)+\lambda_2(v)}{[\s{v-\I(2-\ut)}\s{v+\I(2-\ut)}]^{N/2}},
\label{expLambda0}
\end{equation}
where the terms $\lambda_{1,2}(v)$ are 
\begin{eqnarray}
\lambda_1(v)&:=&\e^{+\beta h/2}\phi(v-\I)\frac{q(v+2\I)}{q(v)},\cr
\lambda_2(v)&:=&\e^{-\beta h/2}\phi(v+\I)\frac{q(v-2\I)}{q(v)},
\end{eqnarray}
and $\phi(v)$ is simply
\begin{equation}
\phi(v):= \left[\s{v-\I(1-\ut)}\s{v+\I(1-\ut)}\right]^{N/2}.
\end{equation}
The function $q(v)$ is defined in terms of the yet to be determined Bethe 
ansatz roots $v_j$
\begin{equation}
q(v):=\prod_j \s{v-v_j}.
\end{equation}
Note that we are mostly interested in $\Lambda$ which is obtained from
$\lambda(v)$ simply by setting $v=0$. Nevertheless, we are led to the study of
the full $v$-dependence since the condition fixing the values of $v_j$ is the
analyticity of $\lambda_1(v)+\lambda_2(v)$ in the complex plane.  This yields
\begin{equation}
a(v_j)=-1,\label{BAQTM}
\end{equation}
where the function $a(v)$ (not to be confused with the Boltzmann weight $a(w)$
above) is defined by
\begin{equation}
\a(v):=\frac{\lambda_1(v)}{\lambda_2(v)}
=\e^{\beta h}\frac{\phi(v-\I)q(v+2\I)}{\phi(v+\I)q(v-2\I)}.
\label{auxfunca}
\end{equation}
Algebraically, we are dealing with a set of coupled non-linear equations
similar to those occurring in the study of the eigenvalues of the
Hamiltonian \cite{Bax82}.  Analytically, there is a profound difference as here
in \refeq{auxfunca} the ratio of $\phi$-functions possesses zeros and poles
converging to the real axis in the limit $N\to\infty$. As a consequence, the
distribution of Bethe ansatz roots is {\it discrete} and shows an {\it
  accumulation point} at the origin, cf.~Fig.~\ref{fig:DistBAR}.  Hence the
treatment of the problem by means of linear integral equations for continuous
density functions \cite{Hul38} is not possible in contrast to the Hamiltonian case.

\section{Manipulation of the Bethe ansatz equations}
\label{BAENLIE}
The eigenvalue expression \refeq{expLambda0} under the subsidiary condition
\refeq{BAQTM} has to be evaluated in the limit $N\to\infty$.  This limit is
difficult to take as an increasing number $N/2$ of Bethe ansatz roots $v_j$
has to be determined. In the Hamiltonian case, i.e.~for \refeq{replacementab}
with $N_1=N$ and $N_2=0$ (or vice versa), the distribution of roots is
continuous and the Bethe ansatz equations \refeq{BAgeneral} can be reduced to
linear integral equations \cite{Hul38,Bax82}. For the general case, this is
no longer possible. In this case, notably for the QTM, the distribution of the
roots is discrete and the standard approach based on continuous density
functions is not possible. From now on we explicitly discuss the QTM case,
i.e.~\refeq{replacementab} with $N_1=N_2=N/2$ yielding all information on the
free energy at arbitrary temperature $T$ including the limit $T=0$. (With some
modifications, the final non-linear integral equations also apply in the case
of the row-to-row transfer matrix and the Hamiltonian. These results, however,
are not of prime interest to this review.)

\subsection{Derivation of non-linear integral equations}
\label{ch:X.2.1}
\begin{figure}[!htb]
  \begin{center}
    \leavevmode
      \includegraphics[width=0.8\textwidth]{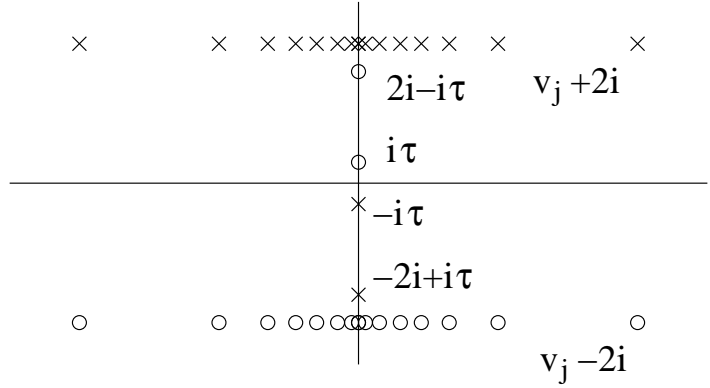}
    \caption{Distribution of zeros $(\circ)$ and poles
      $(\times)$ of the auxiliary function $\a(v)$.
        All zeros and poles $v_j\mp 2\I$ are of first order, the zeros
        and poles at $\pm(2\I-\I \ut)$, $\pm \I \ut$ are of order $N/2$.}
    \label{singa}
  \end{center}
\end{figure}

\begin{figure}[!htb]
  \begin{center}
    \leavevmode
      \includegraphics[width=0.8\textwidth]{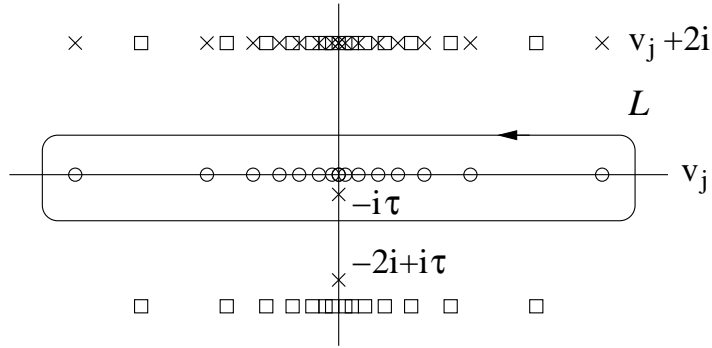}
    \caption{Distribution of zeros and poles of the auxiliary function $\A(v)
      =1+\a(v)$. Note that the positions of zeros $(\circ)$ and poles
      $(\times)$ are directly related to those occurring in the function
      $\a(v)$. There are additional zeros $(\square)$ above and below the real
      axis. The closed contour ${\cal L}$ by definition surrounds the real
      axis and the zeros $(\circ)$ as well as the pole at $-\I \ut$.}
    \label{singA}
  \end{center}
\end{figure}

The main idea of our treatment is the derivation of a set of integral
equations for the function $\a (v)$. This function possesses zeros and poles
related to the Bethe ansatz roots $v_j$, see Fig.~\ref{singa}.
Next we define the associated auxiliary function $\A(v)$ by
\begin{equation}
\A(v)=1+\a(v).
\end{equation}
The poles of $\A(v)$ are identical to those of $\a(v)$. However, the set of
zeros is different. From \refeq{BAQTM} we find that the Bethe ansatz roots are
zeros of $\A(v)$ (depicted by open circles in Fig.~\ref{singA}). There are
additional zeros farther away from the real axis with imaginary parts close to
$\pm 2$. For the sake of completeness, these zeros are depicted in
Fig.~\ref{singA} (open squares), but for a while they are not essential to our
reasoning.
Next we are going to formulate a linear integral
expression for the function $\log\a(v)$ in terms of $\log\A(v)$. 
To this end we consider the function
\begin{equation}
f(v):=\frac 1{2\pi \I}\int_{\cal L}\frac{d}{dv}\log \s{v-w}\log \A(w)dw
\label{f1}
\end{equation}
defined by an integral with closed contour ${\cal L}$ surrounding the real
axis, the parameters $v_j$ and the point $-\I \ut$ in anticlockwise manner,
see Fig.~\ref{singA}.  Note that the number of zeros of $\A(v)$ surrounded by
this contour is $N/2$ and hence is identical to the order of the pole at $-\I
\ut$.  Therefore the integrand $\log\A(w)$ does not show any non-zero winding
number on the contour and consequentially the integral is well-defined. By use
of standard theorems we see that the function $f(v)$ is analytic in the
complex plane away from the real axis (and axes with imaginary parts
tiples of $2\pi/\gamma$). Next we perform an integration by
parts and apply Cauchy's theorem
\begin{equation}
f(v)=\sum_{j}\log \s{v-v_j}-N/2\log \s{v+\I\ut}=
\log\frac{q(v)}{ \s{v+\I\ut}^{N/2}},
\label{f2}
\end{equation}

Thanks to \refeq{f1} and \refeq{f2} we have a linear integral representation
of $\log q(v)$ in terms of $\log\A(v)$. Because of \refeq{auxfunca}
the function $\log\a(v)$ is a linear combination of $\log q$ and explicitly known 
functions leading to
\be
\log \a(v)=\beta h+\log\left(\frac
{ \s{v-\I\ut} \s{v+2\I+\I\ut}}{ \s{v+\I\ut} \s{v+2\I-\I\ut}}\right)^{N/2}
+f(v+2\I)-f(v-2\I).
\label{NLIE1a}
\ee
From now on we use a shorthand notation for the logarithmic derivative of $\s{v}$
\be
\d{v}:=\frac{d}{dv}\log\s{v}=\frac\gamma 2\coth\frac\gamma 2 v.
\ee
By use of the definition of the integration kernel $\varkappa$
\be
\varkappa(u):=\frac1{2\pi\I}\frac{d}{du}\log\frac{\s{u-2\I}}{\s{u+2\I}}
=\frac1{2\pi\I}[\d{u-2\I} - \d{u+2\I}]=\frac\gamma{2\pi}\frac{\sin
2\gamma}{\cosh\gamma u-\cos 2\gamma},
\ee
we may write \refeq{NLIE1a} as
\be
\log \a(v)=\beta h+\frac N2\log\left(\frac
{ \s{v-\I\ut} \s{v+2\I+\I\ut}}{ \s{v+\I\ut} \s{v+2\I-\I\ut}}\right)
-\int_{\cal L}\varkappa(v-w)\log \A(w)dw.
\label{NLIE1}
\ee
This expression for $\a(v)$ is remarkable as it is a non-linear integral
equation (NLIE) of convolution type. It is valid for any value of the Trotter
number $N$ which only enters in the driving (first) term on the
\rhs of \refeq{NLIE1}. This term shows a well-defined limiting behaviour for
$N\to\infty$
\begin{eqnarray}
\frac N2\log\left(\frac
{ \s{v-\I\ut} \s{v+2\I+\I\ut}}{ \s{v+\I\ut} \s{v+2\I-\I\ut}}\right)
&\to& \I N\tau
\left[\frac{d}{dv}\log\s{v+2\I}-\frac{d}{dv}\log\s{v}\right]\cr
&&=\I \beta 2\frac{\sin\gamma}\gamma[\d{v+2\I}-\d{v}],
\end{eqnarray}
leading to a well-defined NLIE for $\a(v)$ even in the limit $N\to\infty$
\begin{equation}
\log\a(v)=\beta h+\beta\epsilon_0(v+\I)
-\int_{\cal L}\varkappa(v-w)\log \A(w)dw,
\label{NLIE2}
\end{equation}
where $\epsilon_0$ is defined by
\be
\epsilon_0(v)=\I[d(v+\I)-d(v-\I)]=2\frac{\sin^2\gamma}{\cosh\gamma v-\cos\gamma}.
\ee
This NLIE allows for a numerical (and in some limiting cases also analytical)
calculation of the function $\a(v)$ on the axes $\Im(v)=\pm 1$.
About the historical development we like to note that NLIEs very similar to
\refeq{NLIE2} were derived for the row-to-row transfer matrix in 
\cite{KluBat90,KluBatPea91}. These equations were then generalized to the
related cases of staggered transfer matrices (QTMs) of the Heisenberg
and RSOS chains \cite{Klu92,Klu93} and the sine-Gordon model \cite{DesVeg92}.

\subsection{Integral expressions for the eigenvalue}
\label{ch:X.2.3a}
In \refeq{NLIE1} and \refeq{NLIE2} we have found integral equations
determining the function $\a$ for finite and infinite Trotter number $N$,
respectively. The remaining problem is the derivation of an expression for
the eigenvalue $\lambda$ in terms of $\a$ or $\A$.

From \refeq{expLambda0} we see that $\lambda(v)$ is an
analytic function with periodicity in imaginary direction (period
$2\pi\I/\gamma$) and exponential asymptotics along the real axis.
Hence, up to a multiplicative constant $C$, we can write
\begin{equation}
\lambda_1(v)+\lambda_2(v)=C\prod_l \s{v-w_l},
\label{expLambda1}
\end{equation}
where $w_l$, $l=1,...,N$, are the zeros of
$\lambda(v)=\lambda_1(v)+\lambda_2(v)$ which are solutions to
$\a(v)=\lambda_1(v)/\lambda_2(v)=-1$, i.e.~ zeros of $\A(v)=1+\a(v)$ that do
not coincide with Bethe ansatz (BA) roots!  These zeros are so-called hole-type solutions to
the BA equations. The holes are located in the complex plane close to the axes
with imaginary parts $\pm 2$, see zeros in Fig.~\ref{singA} depicted by
$\square$.  Thanks to Cauchy's theorem we find for $v$ in the neighbourhood of
the real axis
\begin{equation}
\frac 1{2\pi \I}\int_{\cal L}\d{v-w-2\I}[\log \A(w)]'dw
=\sum_j\d{v-v_j-2\I}-\frac N2\d{v+\I\ut-2\I}
\label{IE1}
\end{equation}
as the only singularities of the integrand surrounded by the contour ${\cal
  L}$ are the simple zeros $v_j$ and the pole $-\I\ut$ of order $N/2$ of the
function $\A$.  Also, for $v$ in the neighbourhood of the real axis we obtain
\bea
\frac 1{2\pi \I}\int_{\cal L}\d{v-w}[\log \A(w)]'dw
&=&\sum_j\d{v-v_j-2\I}-\sum_l\d{v-w_l}\cr
&&+\frac N2\d{v+2\I-\I\ut},
\label{IE2}
\eea
where the evaluation of the integral has been done by use of the singularities
outside of the contour ${\cal L}$ and use of the period $2\pi\I/\gamma$ of the
integrand. The old contour ${\cal L}$ is replaced by a contour
${\widetilde{\cal L}}$ such that the upper (lower) part of
${\widetilde{\cal L}}$ is the lower part of ${\cal L}$ (the upper part of
${\cal L}$ shifted by $-2\pi\I/\gamma$), and reversed orientation. The surrounded
singularities are the simple poles $v_j+2\I-2\pi\I/\gamma$, the zeros $w_l$
with or without shift $-2\pi\I/\gamma$, and the pole $\I\ut-2\I$ of order
$N/2$ of the function $\A$.

Next, we take the difference of \refeq{IE1} and \refeq{IE2}, perform an
integration by parts with respect to $w$, and finally integrate with
respect to $v$
\bea
&&\frac 1{2\pi \I}\int_{\cal L}\left[\d{v-w}-\d{v-w-2\I}\right]\log \A(w)dw\cr
&&=\log\frac{[\s{v-\I(2-\ut)}\s{v+\I(2-\ut)}]^{N/2}}{\prod_l\s{v-w_l}}+\hbox{const.}
\label{IE3}
\eea
Combining
\refeq{expLambda1}, \refeq{IE3} and \refeq{expLambda0} we find 
\begin{equation}
\log\lambda(v)=-\beta h/2-
\frac 1{2\pi \I}\int_{\cal L}\left[\d{v-w}-\d{v-w-2\I}\right]\log \A(w)dw,
\label{expLambda2}
\end{equation}
where the constant was determined from the asymptotic behaviour for
$v\to\infty$ and use of $\lambda(\infty)=\exp(\beta h/2)+\exp(-\beta h/2)$
and $\A(\infty)=1+\exp(\beta h)$.

These formulas, \refeq{expLambda2} and \refeq{NLIE2}, are the basis of an
efficient analytical and numerical treatment of the thermodynamics of the
Heisenberg chain. There are, however, variants of these integral equations
that are somewhat more convenient for the analysis, especially for magnetic
fields close to 0. The alternative integral expression for $\Lambda$ reads
\cite{Klu92,Klu93}
\begin{align}
&\ln\Lambda=-\beta e_0
+\int_{-\infty}^{\infty}K(x)\ln[\mfA(x)\mfAb(x)]dx,
&K(x)=\frac{1}{4\cosh{\frac{\pi}{2}x}},
\label{largestalt}
\end{align}
where $\mfA(x)$ and $\mfAb(x)$ are complex-valued functions with integration
paths along the real axis. These functions are determined from the following
set of non-linear integral equations
\begin{align}
&\ln\mfa(x)=-\beta\frac{\sin\gamma}\gamma\frac{\pi}{\ch\frac{\pi}{2}x}+\frac{\pi\beta
            h}{2(\pi-\gamma)}+
            \kappa\ast\ln\mfA(x)-\kappa\ast\ln\mfAb(x+2i),\label{nliealt1}\db
&\ln\mfab(x)=-\beta\frac{\sin\gamma}\gamma\frac{\pi}{\ch\frac{\pi}{2}x}-\frac{\pi\beta
            h}{2(\pi-\gamma)}+
            \kappa\ast\ln\mfAb(x)-\kappa\ast\ln\mfA(x-2i),\label{nliealt2}\db
&\mfA(x)=1+\mfa(x),\qquad \mfAb(x)=1+\mfab(x).
\label{nliealt}
\end{align}
The symbol $\ast$ denotes the convolution $f\ast g(x)=
\int_{-\infty}^{\infty} f(x-y)g(y)dy$ and the function 
$\kappa(x)$ is defined by
\be
\kappa(x)=\frac{1}{2\pi}\int_{-\infty}^{\infty}
          \frac{\sh\left(\frac{\pi}{\gamma}-2\right)k}
                {2\ch k \sh\left(\frac{\pi}{\gamma}-1\right)k} \e^{\i kx}dk.
\label{bulk}
\ee 
Note that the integrals in \refeq{nliealt1} and \refeq{nliealt2} are
well-defined with integration paths just below and above the real axis.

The above equations are obtained from \refeq{NLIE2} by a partial
``particle-hole'' transformation of the function $a(v)$ only on the axis
$\Im(v)=- 1$.  Replacing $\log\A=\log\bA+\log a$ (where $\ba=1/a$, $\bA=1+\ba$) on the
lower part of ${\cal L}$ in \refeq{NLIE2} leads to an equation involving
convolution type integrals with $\log\A$, $\log\bA$ and $\log a$. This
equation can be resolved explicitly for $\log a$ by straightforward
calculations in ``momentum space''. Finally, $\mfa(x):=a(x+\I)$ and $\mfab(x):=\ba(x-\I)$.

\section{Numerical results for thermodynamical quantities}
\label{Numer}
By numerical integration and iteration the integral equation (\ref{NLIE2}) can
be solved on the axes $\Im(v)=\pm 1$ defining functions
$\a^\pm(x):=\a(x\pm\I)$. Alternatively and particularly convenient for the case of
vanishing magnetic field $h$, equations \refeq{nliealt1} and \refeq{nliealt2}
can be used for the functions $\mfa$ and $\mfab$. Choosing appropriate initial
functions the series $\a^\pm_{k}$ with $k=0,1,2,\dots$ converges rapidly.  In
practice only a few steps are necessary to reach high-precision results.
Moreover, using the well-known Fast Fourier Transform algorithm we can compute
the convolutions very efficiently. In fact, some of the convolutions in
(\ref{NLIE2}) or (\ref{nliealt1},\ref{nliealt2}) are delicate to be evaluated
in ``real space'', because of the appearance of a pole of the kernel just at
the integration contour. These integrals are automatically handled correctly
in ``momentum space''.

In order to calculate derivatives of the thermodynamical potential with
respect to temperature $T$ and magnetic field $h$ one
can avoid  numerical differentiations  by  utilizing similar  integral
equations guaranteeing the  same  numerical accuracy as for  the  free
energy. The idea is as follows. Consider the function 
\begin{equation*}
l\a_{\beta}:=
\frac{\partial}{\partial\beta}\log\a
\quad\text{with}\quad
\frac{\partial}{\partial\beta}\log(1+\a)=
\frac{1}{1+\a}\frac{\partial\a}{\partial\beta}=
\frac{\a}{1+\a}\,l\a_{\beta},
\end{equation*}
\begin{figure}[!htb]
  \begin{center}
    \includegraphics[width=0.7\textwidth]{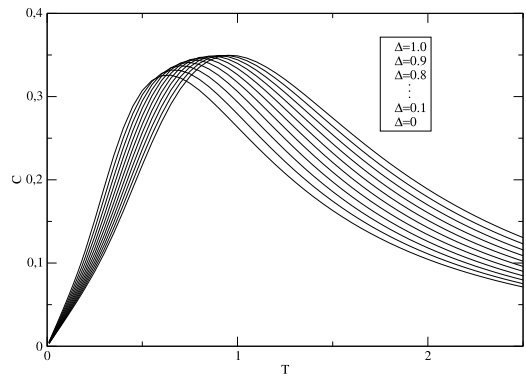}
    \caption{Specific heat $c(T)$ data versus temperature $T$ for the spin-1/2
      $XXZ$ chain with repulsive interaction $0\le\Delta\le 1$.}
    \label{Fig3a}
  \end{center}
\end{figure}
we directly obtain from \refeq{NLIE2} a {\it linear} integral equation for
$l\a_{\beta}$ if we regard the function $\a$ as given. Once the integral
equation \refeq{NLIE2} is solved for $\a$, the integral equation for
$l\a_{\beta}$ associated with \refeq{NLIE2} can be solved. In this manner, we
may continue to any order of derivatives with respect to $T$ (and $h$).
However, in practice only the first and second orders matter. Here we restrict
our treatment to the specific heat $c(T)$ and the magnetic susceptibility
$\chi(T)$ (derivatives of second order with respect to $T$ and $h$), see
Figs.~\ref{Fig3a} and \ref{Fig3b}.

\begin{figure}[!htb]
  \begin{center}
    \includegraphics[width=0.7\textwidth]{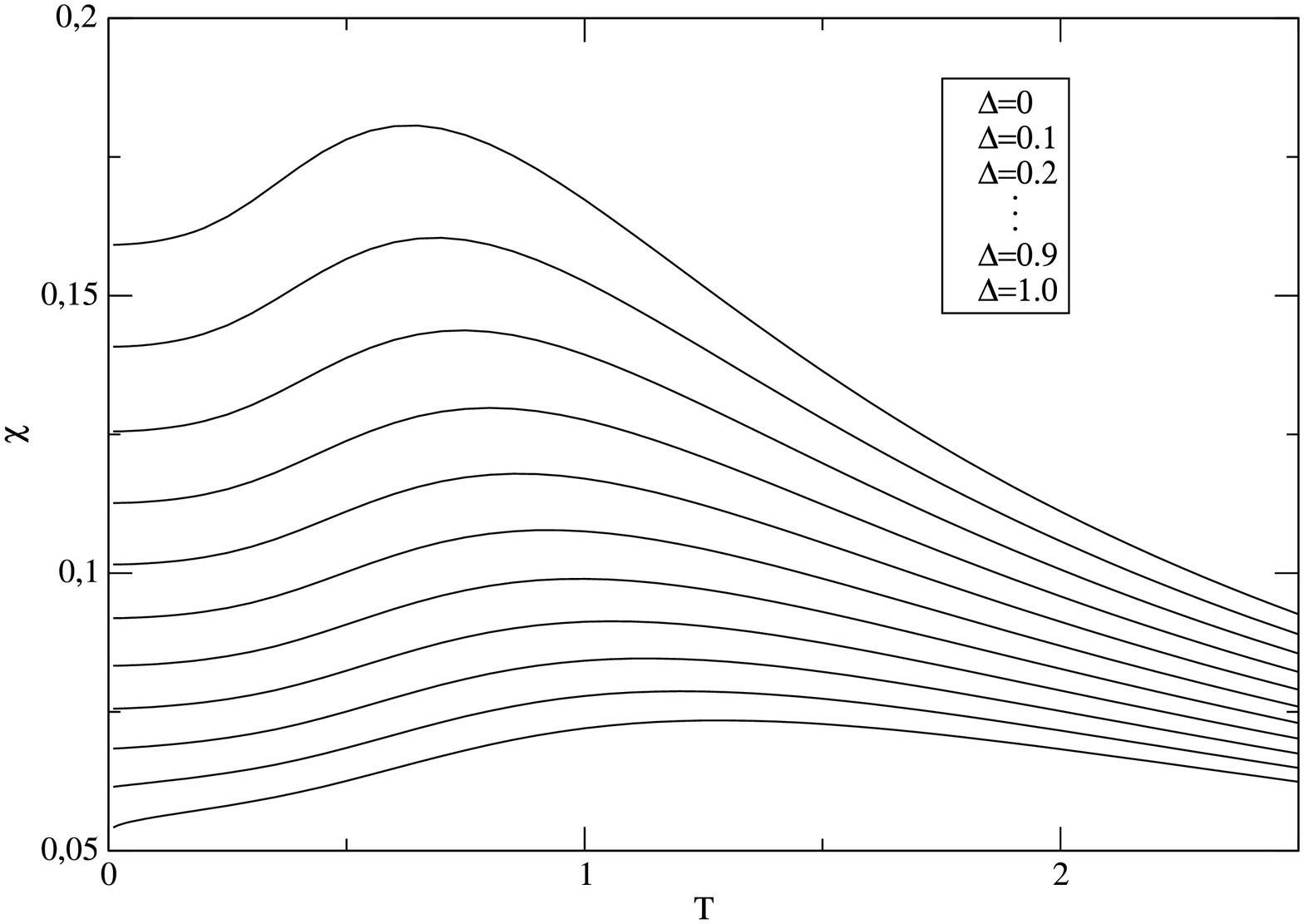}
    \caption{Susceptibility $\chi(T)$ data versus temperature $T$ for the
      spin-1/2 $XXZ$ chain with repulsive interaction $0\le\Delta\le 1$.}
    \label{Fig3b}
  \end{center}
\end{figure}
Note the characteristic behaviour of $c(T)$ and $\chi(T)$ at low temperatures.
The linear behaviour of $c(T)$ and the finite ground-state limit of $\chi(T)$
are manifestations of the linear energy-momentum dispersion of the low-lying
excitations (spinons) of the isotropic antiferromagnetic Heisenberg chain.
Also, with increase of the repulsion $\Delta$ the location of the finite
temperature maximum in $c(T)$ shifts to higher temperatures and the values of
$\chi(T)$ drop.  In the high temperature limit the asymptotics of $c(T)$ and
$\chi(T)$ are $1/T^2$ and $1/T$. This and the existence of the finite
temperature maximum are a consequence of the finite dimensional local degree of
freedom, i.e.~the spin per lattice site.

\begin{figure}[!htb]
  \begin{center}
    \includegraphics[width=0.8\textwidth]{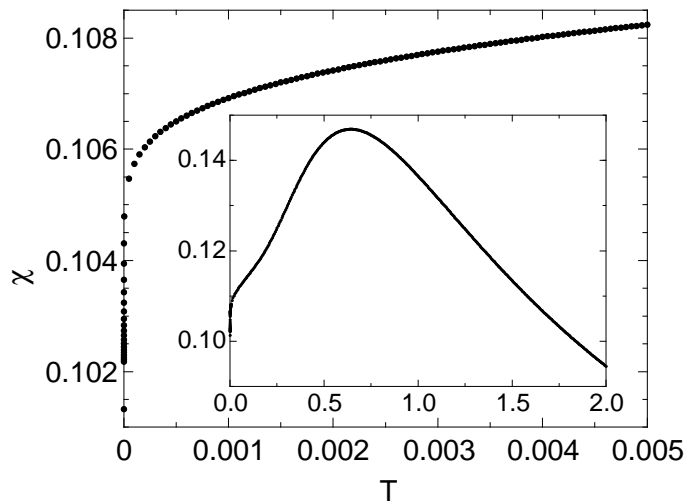}
    \caption{Magnetic susceptibility $\chi$ at low temperature $T$ for the
      isotropic spin-1/2 $XXX$ chain. In the inset $\chi(T)$ is shown on a
      larger temperature scale.}
    \label{Fig1}
  \end{center}
\end{figure}

For $\Delta=1$ note that $\chi(T)$ approaches the ground-state limit $\chi(0)$
in a singular manner, see also Fig.~\ref{Fig1}.  The numerical data at
extremely low temperatures provide evidence of logarithmic correction terms,
see also \cite{Eggert1994} and later lattice studies
\cite{Klumper1998,KlJohn00} confirming the field theoretical treatment by
\cite{Lukyanov1998}. These logarithmic terms are responsible for the infinite
slope of $\chi(T)$ at $T=0$ despite the finite ground-state value
$\chi(0)=1/\pi^2$. Precursors of such strong slopes have been seen in
experiments down to relatively low temperatures, see
e.g.~\cite{Takagi96}. Unfortunately, most quasi one-dimensional quantum spin
systems undergo a phase transition at sufficiently low temperatures driven by
residual higher dimensional interactions. Hence the onset of quantum critical
phenomena of the Heisenberg chain at $T=0$ becomes visible, but cannot be
identified beyond all doubts.

\section{Thermal transport}
\label{transport}
The simplest approach to the investigation of transport properties is based on
linear response theory leading to the Kubo formulas \cite{kubo,Mahan} relating
conductivities to dynamical correlation functions of local current
operators. Hence, the calculation of transport properties is more difficult
than the computation of the free energy. In fact, the calculation of
correlation functions is a most difficult issue and exact results are rare,
even for integrable systems.  In some situations, however, the explicit
computation of correlation functions can be avoided. For the spin transport of
the Heisenberg chain ($\equiv$ electrical transport of the model in the
particle representation) the Drude weight of the dynamical conductivity can be
cast into a form involving only spectral properties without explicit recourse
to matrix elements. At the time of writing, these expressions are known how to
be evaluated for zero temperature, the case of non-zero temperature is still
controversial.

A fortuitous case is the thermal transport as the thermal current ${\mathcal
  J}_{\rm E}$ is one of the conserved currents $\mathcal{J}^{(n)}$
  \refeq{generator}. The first three conserved currents $(n=0, 1, 2)$ are
  related to the momentum operator, the Hamiltonian and the thermal current via
\begin{align}
&P=-\i\mathcal{J}^{(0)}\\
&H=2\frac{\sin\gamma}{\gamma}\mathcal{J}^{(1)}
             -\frac{L}{2} \Delta,\nn \\
&\mathcal{J}_{\rm E}=
             \i \left(2\frac{\sin\gamma}{\gamma}\right)^2
              \mathcal{J}^{(2)}+
              \i L.
\label{conserve} 
\end{align}
Note that the spin current (electrical current) is not conserved as it is not
contained in the sequence of conserved currents $\mathcal{J}^{(n)}$!

Let us first motivate that $\mathcal{J}^{(2)}$ is indeed the thermal current
$\mathcal{J}_{\rm E}$ of the system. To this end we impose the continuity
equation relating the time derivative of the local Hamiltonian (interaction)
and the divergence of the current: $\dot h=-{\rm div}\, j^{\rm E}$.  The time
evolution of the local Hamiltonian $h_{k,k+1}$ is obtained from the commutator
with the total Hamiltonian and the divergence of the local current on the
lattice is given by a difference expression
\be
\frac{\partial h_{k k+1}(t)}{\partial t}
=\i[H,h_{k k+1}(t)]=-\{j_{k+1}^{\rm E}(t)-j_{k}^{\rm E}(t)\}.
\ee
Apparently the last equation is satisfied 
with a local thermal current operator $j^{\rm E}_{k}$
defined by
\be
j_k^{\rm E}=\i[h_{k-1 k},h_{k k+1}].
\ee
In fact, up to a trivial scale and shift, the conserved current
$\mathcal{J}^{(2)}$ is identical to the \rhs of the upper equation, see also
\cite{znp,lu,gm,Tsv,Frahm,Racz}.

The Kubo formulas \cite{kubo,Mahan} are obtained within linear response theory
and yield the (thermal) conductivity $\kappa$ relating the (thermal) current
${\mathcal J}_{\rm E}$ to the (temperature) gradient $\nabla T$
\be
{\mathcal J}_{\rm E}=\kappa \nabla T, 
\ee 
where
\be
\kappa(\omega)=\beta\int_0^\infty dt \e^{-\i\omega t}\int_0^\beta
d\tau \langle {\mathcal J}_{\rm E}(-t-\i\tau) {\mathcal J}_{\rm
E}\rangle.
\ee
As the total thermal current operator ${\mathcal J}_{\rm E}$ commutes with the
Hamiltonian $H$ of the $XXZ$ chain we find
\be
\kappa(\omega)=\frac 1{\i(\omega-\i\epsilon)}\beta^2
\langle{\mathcal J}_{\rm E}^2\rangle, \qquad (\epsilon\to0+),
\ee 
with $\Re\kappa(\omega)=\tilde\kappa\delta(\omega)$ 
where
\be
\tilde\kappa=\pi\beta^2\langle{\mathcal J}_{\rm E}^2\rangle.
\ee
As a consequence, the thermal conductivity at zero frequency is infinite! This
is only natural, as the conserved current cannot decay in time. However, the weight of
the zero-frequency peak is some finite and non-trivial temperature dependent
quantity to be calculated from the second moment of the thermal
current. 

Quite generally, the expectation values of conserved quantities may be
calculated by use of a suitable generating function
\be
Z={\rm Tr}\,\exp(-\beta\mathcal{H}+\lambda \mathcal{J}_{\rm E}),
\label{part}
\ee
from which we find the expectation values by derivatives with respect to 
$\lambda$ at $\lambda=0$
\be
\frac{\partial}{\partial \lambda}\ln Z \Big|_{\lambda=0}
=\langle \mathcal{J}_{\rm E} \rangle=0,\qquad
\left(\frac{\partial}{\partial \lambda}\right)^2\ln Z \Big|_{\lambda=0}
=\langle \mathcal{J}_{\rm E}^2 \rangle
-\langle \mathcal{J}_{\rm E} \rangle^2
=\langle \mathcal{J}_{\rm E}^2 \rangle,
\ee
where we have used that the expectation value of the thermal current
in thermodynamical equilibrium is zero. 

Instead of $Z$ we will find it slightly more convenient to work with 
a partition function
\be
\Z={\rm Tr}\,\exp(-\lambda_1\mathcal{J}^{(1)}-\lambda_n\mathcal{J}^{(n)}),
\label{part2}
\ee
notably $n=2$. With view to \refeq{conserve} we choose
\be
\lambda_1=\beta \left(2\frac{\sin\gamma}{\gamma}\right),\qquad
\lambda_2=-\i \lambda \left(2\frac{\sin\gamma}{\gamma}\right)^2,
\ee
and obtain the desired expectation values from $\Z$
\be
\langle \mathcal{J}_{\rm E}^2 \rangle=
\left(\frac{\partial}{\partial \lambda}\right)^2\ln\Z \Big|_{\lambda=0}.
\ee

We can deal with $\Z$ rather easily. Consider the trace of a product 
of $N$ row-to-row transfer matrices $T(u_j)$ with some spectral parameters
$u_j$ close to zero, but still to be specified, and the $N$th power of
the inverse of $T(0)$
\bea
Z_N&=&{\rm Tr}\,\left[T(u_1)\cdot\, \dotsb \,\cdot T(u_N)\cdot T(0)^{-N}\right]\cr
&=&{\rm Tr}\,\exp\left(\sum_j[\ln T(u_j)-\ln T(0)]\right).
\eea
Now it is a standard exercise in arithmetic to devise a sequence of
$N$ numbers $u_1$,...,$u_N$ (actually $u_j=u_j^{(N)}$) such that 
\be
\lim_{N\to\infty}\sum_j[f(u_j)-f(0)]=
-\lambda_1\frac{\partial}{\partial u}f(u)\Big|_{u=0}
-\lambda_n\left(\frac{\partial}{\partial u}\right)^nf(u)\Big|_{u=0}.
\ee
We only need the existence of such a sequence of numbers, the precise
values are actually not important. In the limit $N\to\infty$ we note
\be
\lim_{N\to\infty}Z_N=\Z.
\ee

We can proceed along the established path of the quantum transfer matrix (QTM)
formalism presented above and derive the partition function $Z_N$ in the thermodynamic limit
$L\to\infty$
\be
\lim_{L\to\infty}Z_N^{1/L}=\Lambda,
\ee
where $\Lambda$ is the largest eigenvalue of the QTM. The
integral expression for $\Lambda$ reads
\begin{align}
&\ln\Lambda=\sum_j[e(u_j)-e(0)]
+\int_{-\infty}^{\infty}K(x)\ln[\mfA(x)\mfAb(x)]dx,
\label{largest}
\end{align}
with $K(x)$ already defined in \refeq{largestalt} and
some function $e(x)$ given in \cite{Klu92,Klu93}. In the limit $N\to\infty$
the first term on the \rhs of the last equation turns into
\be
\lim_{N\to\infty}\sum_j[e(u_j)-e(0)]=
-\lambda_1\frac{\partial}{\partial u}e(u)\Big|_{u=0}
-\lambda_n\left(\frac{\partial}{\partial u}\right)^ne(u)\Big|_{u=0},
\ee
a rather irrelevant term as it is linear in $\lambda_1$ and $\lambda_n$,
and therefore the second derivatives with respect
to $\lambda_1$ and $\lambda_n$ vanish.
The functions $\mfA(x)$ and $\mfAb(x)$ are determined from
the following set of non-linear integral equations
\begin{align}
&\ln\mfa(x)=\sum_j[\varepsilon_0(x-\i u_j)-\varepsilon_0(0)]+
            \kappa\ast\ln\mfA(x)-\kappa\ast\ln\mfAb(x+2i),\nn \db
&\ln\mfab(x)=\sum_j[\varepsilon_0(x-\i u_j)-\varepsilon_0(0)]+
            \kappa\ast\ln\mfAb(x)-\kappa\ast\ln\mfA(x-2i),\nn \db
&\mfA(x)=1+\mfa(x),\qquad \mfAb(x)=1+\mfab(x).
\label{nlie}
\end{align}
with a function $\varepsilon_0(x)$ given in terms of hyperbolic
functions \cite{Klu92,Klu93}.
Again, the summations in \refeq{nlie} can be simplified in the
limit $N\to\infty$
\be
\lim_{N\to\infty}\sum_j[\varepsilon_0(x-\i u_j)-\varepsilon_0(x)]=
-\lambda_1\underbrace{
\left(-\i\frac{\partial}{\partial x}\right)\varepsilon_0(x)
}_{=:\varepsilon_1(x)}
-\lambda_n\underbrace{\left(-\i\frac{\partial}{\partial x}
\right)^n\varepsilon_0(x)}_{=:\varepsilon_n(x)}.
\label{drivter}
\ee
where the first function can be found in \cite{Klu92,Klu93} and is simply
\be
\varepsilon_1(x)=2\pi K(x)=
\frac{\pi}{2\ch\frac{\pi}{2}x},
\ee
and hence the second function is
\be
\varepsilon_n(x)=\left(-\i\frac{\partial}{\partial
x}\right)^{n-1}\varepsilon_1(x).
\ee
We like to note that the structure of the driving term \refeq{drivter}
appearing in the NLIE \refeq{nlie} reflects the structure of the
generalized Hamiltonian in the exponent on the \rhs of \refeq{part2}. 
We could have given an alternative derivation
of the NLIE along the lines of the thermodynamic Bethe ansatz (TBA). 
In such an approach the driving term is typically the one-particle energy 
corresponding to the generalized Hamiltonian. Hence it has contributions
due to the first as well as the $n$th logarithmic derivative of the
row-to-row transfer matrix, i.e.~the terms $\varepsilon_1$ and $\varepsilon_n$.

\begin{figure}
\begin{center}
\includegraphics[width=\textwidth]{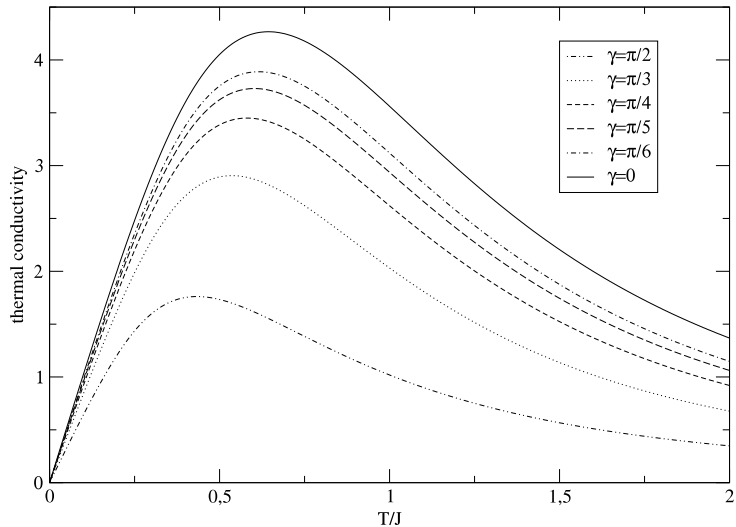}
\end{center}
\caption{Illustration of numerical results for the thermal conductivity
$\tilde\kappa$ as a function of temperature $T$ for various anisotropy
parameters $\Delta=\cos\gamma$ with $\gamma=0, \pi/6, \pi/5, \pi/4, \pi/3,
\pi/2$.  }
\label{kappa}
\end{figure}

In Fig.~\ref{kappa} we show $\tilde\kappa(T)$ for various anisotropy
parameters $\gamma$. Note that $\tilde\kappa(T)$ has linear $T$ dependence
at low temperatures. 
At high temperatures $\tilde\kappa(T)$ behaves like $1/T^2$. 

\section{Summary}
%
We have reviewed the treatment of integrable quantum systems at zero and
finite temperature based on a lattice path integral formulation and the
definition of the so-called quantum transfer matrix (QTM). In detail, the
thermodynamical properties of the partially anisotropic Heisenberg chain were
discussed. As we hope, a transparent analysis of the eigenvalue problem of the
row-to-row and quantum transfer matrices has been given, resulting in a set of
non-linear integral equations (NLIE). From a numerical solution of these NLIEs
at arbitrary temperature the specific heat and magnetic susceptibility data
were obtained. Also the Drude weight of the thermal current was calculated.
Although we have given the explicit results only for the case of vanishing
magnetic field ($h=0$), the derivation of the free energy is not restricted to
this case. In fact, the NLIEs given above are valid even in the case of
non-vanishing fields $h$. However, the above treatment of the thermal
conductivity is limited to the case $h=0$. The generalization to arbitrary
fields is an open problem as is practically any physical question involving
the explicit knowledge of correlation functions at zero and even more at
finite temperature. 

The purpose of writing this review was to convince the reader that research on
integrable quantum systems is an attractive and rewarding field of science.
Hopefully, this goal has been reached and the reader has gained from this
presentation enough insight to begin or continue his or her own
investigations.

%
\section*{Acknowledgments}
The author acknowledge financial support by the Deutsche
Forschungsgemeinschaft under grant No.~645/4-1 and the
Schwerpunktprogramm SP1073.
%
%

\bibliographystyle{phys}
\bibliography{../Book/thermo,../Book/hub}

\sloppy

\printindex
\end{document}